\documentclass[aps,pra,twocolumn,showpacs,superscriptaddress]{revtex4-1}

\usepackage{graphicx,color}
\usepackage{amsfonts}
\usepackage{amsmath}
\usepackage{amssymb}

\begin{document}
\title{Optimal preparation of graph states}
\author{Ad\'{a}n Cabello}
 \email{adan@us.es}
 \affiliation{Departamento de F\'{\i}sica Aplicada II,
 Universidad de Sevilla, E-41012 Sevilla, Spain}
\author{Lars Eirik Danielsen}
 \email{larsed@ii.uib.no}
 \affiliation{Department of Informatics,
 University of Bergen, P.O. Box 7803, Bergen N-5020, Norway}
\author{Antonio J. L\'{o}pez-Tarrida}
 \email{tarrida@us.es}
 \affiliation{Departamento de F\'{\i}sica Aplicada II,
 Universidad de Sevilla, E-41012 Sevilla, Spain}
\author{Jos\'{e} R. Portillo}
 \email{josera@us.es}
 \affiliation{Departamento de Matem\'{a}tica Aplicada I,
 Universidad de Sevilla, E-41012 Sevilla, Spain}
\date{\today}




\begin{abstract}
We show how to prepare any graph state of up to 12 qubits with:
(a) the minimum number of controlled-$Z$ gates, and (b) the
minimum preparation depth. We assume only one-qubit and
controlled-$Z$ gates. The method exploits the fact that any
graph state belongs to an equivalence class under local
Clifford operations. We extend up to 12 qubits the
classification of graph states according to their entanglement
properties, and identify each class using only a reduced set of
invariants. For any state, we provide a circuit with both
properties (a) and (b), if it does exist, or, if it does not,
one circuit with property (a) and one with property (b),
including the explicit one-qubit gates needed.
\end{abstract}


\maketitle


\section{Introduction}
\label{Sec1}


Building a quantum computer entails isolating atomic-scale
systems, except when a controlled interaction (e.g., a logic
gate) is applied. Isolation must be kept up until a subsequent
controlled interaction is applied. Any undesirable coupling
with the outside disrupts the quantum state of the systems and
ruins the computation. Due to the enormous difficulties which
keeping atomic-scale systems isolated and controlled entails, a
main limiting factor preventing the development of quantum
computing is the amount of time during which the systems must
be kept isolated and controlled to achieve the computation.
Therefore, the task of reducing this amount of time is an
essential factor in the process of achieving the goal of
building a quantum computer.

Another limiting factor for quantum computation is the number
of entangling gates needed. While one-qubit gates can be built
with fidelities higher than $99\%$, two-qubit entangling gates
hardly reach $93\%$, and this becomes worse for three-qubit
gates, etc. Therefore, since one- and two-qubit gates are
enough for universal quantum computation, it is reasonable to
focus on circuits with only one- and two-qubit gates, and
having the least possible number of two-qubit gates.

Here we address the problem of preparating an important class
of quantum states with circuits requiring (a) the minimum
number of two-qubit entangling gates, and (b) the minimum
preparation depth (i.e., requiring a minimum number of time
steps). We assume that we can implement arbitrary one-qubit
gates and one specific two-qubit entangling gate, the
controlled-$Z$ gate. The result can be easily extended to any
other specific two-qubit gate.


\subsection{Graph states}

Graph states \cite{HEB04, HDERVB06} are a type of $n$-qubit
pure state with multiple applications in quantum information
processing. Two important examples are: in quantum
error-correction, the stabilizer codes which protect quantum
systems from errors \cite{Gottesman96, SW02, Schlingemann02}
and, in measurement-based (or one-way) quantum computation
\cite{RB01}, the initial states consumed during the computation
\cite{VMDB06}.

The definition of a graph state already provides a blueprint
for its preparation: an $n$-qubit graph state $|G\rangle$ is a
pure state associated with a graph $G=(V, E)$ consisting of a
set of $n$ vertices $V=\{0,\ldots,n-1\}$ and a set of edges $E$
connecting pairs of vertices, $E \subset V \times V$
\cite{HEB04, HDERVB06}. Each vertex represents a qubit. An edge
$(i,j) \in E$ represents an Ising-type interaction between
qubits $i$ and $j$. To prepare $|G\rangle$, first prepare each
qubit in the state
$|+\rangle=\frac{1}{\sqrt{2}}(|0\rangle+|1\rangle)$, i.e., the
initial state will be $|\psi_0\rangle=\bigotimes_{i\in
V}|+\rangle_{i}$. Then, for each edge $(i,j) \in E$ connecting
two qubits $i$ and $j$, apply a controlled-$Z$ gate between
these qubits, i.e., the unitary transformation
$C_Z=|00\rangle\langle00|+|01\rangle\langle01|+|10\rangle\langle10|-|11\rangle\langle11|$.

\subsection{Preparation using only controlled-$Z$ gates}

\begin{figure}[t]
\includegraphics[width=0.22\textwidth]{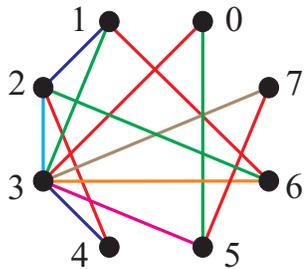}\\
\caption{\label{grafo-complex8}Graph corresponding to the 8-qubit graph state we want to prepare.
Edges in the same color represent controlled-$Z$ gates that can be performed in the same time step.}
\end{figure}

Let us suppose that we have the state
$|\psi_0\rangle=\bigotimes_{i=0}^{n-1}|+\rangle_i$, and we want
to prepare the eight-qubit graph state $|G\rangle$ whose graph
$G$ is in Fig.~\ref{grafo-complex8}, using only controlled-$Z$
gates. One possible way is to follow the preparation procedure
suggested by $G$, taking into account the possible restrictions
in performing two or more controlled-$Z$ gates simultaneously,
and optimally distributing the controlled-$Z$ gates to minimize
the number of steps. The {\em preparation depth} of $|G\rangle$
is the {\em minimum} number of time steps required to prepare
$|G\rangle$ \cite{MP04}.


The state $|\psi_0\rangle$ corresponds to a graph with $n$
isolated vertices. The total number of edges in $G$ gives a
trivial upper bound to the preparation depth of $|G\rangle$,
since each controlled-$Z$ can be applied in a time step. To
find the minimum number of time steps, we have to explore the
possibility of applying two or more controlled-$Z$ gates in a
single time step.

Given a vertex $i$ in $G$, the {\em neighborhood} of $i$,
${\mathcal N}(i)$, is the set of vertices connected to $i$. Now
let us suppose that there is more than one element in
$\mathcal{N}(i)$, i.e., $|\mathcal{N}(i)|>1$, and let $j,k \in
\mathcal{N}(i)$ be two of the neighbors. Then, to prepare the
corresponding graph state $|G\rangle$ we must apply a
controlled-$Z$ gate to entangle qubits $i$ and $j$, and another
one to entangle qubits $i$ and $k$. Applying a controlled-$Z$
gate between qubits $i$ and $j$ in a certain time step implies
that both qubits are busy during this entangling interaction.
If we focus on qubit $i$, we have to wait for a further time
step to have qubit $i$ free before applying another
controlled-$Z$ gate to entangle $i$ and $k$. Nevertheless,
vertex $k$ could be connected to another vertex $l \neq \{i,
j\}$ in $G$. If such is the case, we could in principle take
advantage of the same time step we are using to entangle qubits
$i$ and $j$ in order to do the same with qubits $k$ and $l$,
because both controlled-$Z$ gates can be performed in parallel.

Remarkably, the problem of determining the {\em minimum} number
of time steps with the restrictions that we have mentioned is
related to an old problem in graph theory: given an edge
$(i,j)$ in $G$, let us use a certain color to mark $(i,j)$ and
those other edges of $G$ corresponding to controlled-$Z$ gates
that can be performed at the same time step than that of
$(i,j)$, and use different colors for those edges related to
controlled-$Z$ gates that do not fulfill this condition. Since
two controlled-$Z$ gates can be performed at the same time step
if and only if the four qubits involved do not coincide, then
every edge incident to the same vertex must have a {\em
different} color. In graph theory this color configuration is
called a {\em proper edge coloring} or, to put it more
concisely, the graph is said to be edge-colored. Hence, the
preparation depth problem is equivalent to determining the
minimum number of colors required to get a proper edge coloring
of $G$. This problem is known as the determination of the {\em
chromatic index} or {\em edge chromatic number} $\chi'(G)$.

Let us denote by $\Delta(G)$ the maximum degree of $G$ (i.e.,
the maximum number of edges incident to the same vertex).
Vizing's theorem \cite{Vizing64} states that $G$ can be
edge-colored in either $\Delta(G)$ or $\Delta(G)+1$ colors, and
not fewer than that. Therefore, $\chi'(G)$ is either
$\Delta(G)$ or $\Delta(G)+1$. A proof can be found in
\cite{MG92}. The important point is that $\chi'(G)$ gives the
preparation depth of $|G\rangle$ \cite{MP04, BK06}.

Graphs requiring $\Delta(G)$ colors are called class-$1$
graphs, and those requiring $\Delta(G)+1$ colors are called
class-$2$ graphs. For instance, the four-qubit fully connected
graph state is represented by a graph of four vertices, six
edges, and maximum degree equal to 3: it is a class-1 graph, so
that its preparation depth coincides with its maximum degree:
3. On the other hand, the three-qubit fully connected graph
state is represented by a graph of three vertices with three
edges and maximum degree equal to 2: it is the smallest class-2
graph and, as a consequence, its preparation depth is also 3.

\begin{figure}[t]
\includegraphics[width=0.45\textwidth]{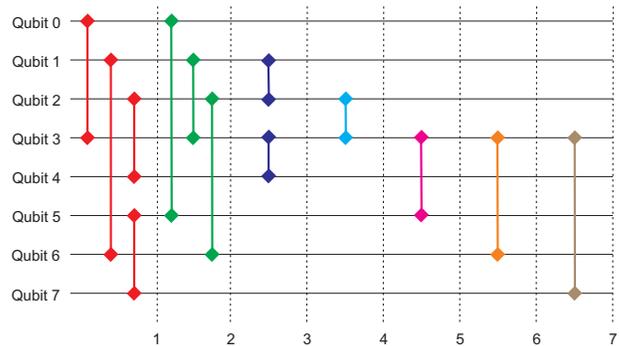}\\
\caption{\label{prep-complex8} A circuit with minimum preparation depth for the graph state
corresponding to Fig.~\ref{grafo-complex8}, using only controlled-$Z$
gates. In the circuit, a qubit (vertex of the graph) is
represented by a horizontal wire, and a controlled-$Z$ gate
(edge in the graph) by a vertical segment with diamond-shaped
ends connecting the qubits involved in the operation. Time
steps are separated by vertical dashed lines.}
\end{figure}

The graph in Fig.~\ref{grafo-complex8} is a class-1 graph: it
can be edge-colored with $\Delta(G)=7$ colors (this number
corresponds to the degree of vertex~3). Hence, if we use only
controlled-$Z$ gates, the minimum preparation depth of the
corresponding graph state is 7. For instance, one of the
optimal distributions of the 13 controlled-$Z$ gates is given
in the circuit of Fig.~\ref{prep-complex8}, which has seven
time steps.


However, if we are allowed to use one-qubit gates, in most
cases it is possible to get the desired graph state with a
lower number of two-qubit gates and less preparation depth. For
that purpose, one has to take into account the degree of
entanglement of the state we want to prepare.

\subsection{Optimum preparation}

Two graph states have the same degree of entanglement if and
only if they are equivalent under local unitary (LU) operations
\cite{VDM04x}, so that they belong to the same LU-equivalence
class. Moreover, previous results suggest that, for graph
states of up to 26 qubits \cite{JCWY07}, the notions of
LU-equivalence and LC-equivalence (equivalence under local
Clifford operations) coincide and, therefore, entanglement
classes are in fact local Clifford equivalence classes (LC
classes). This implies a remarkable simplification, since it is
possible to carry out a graphical description of the action of
local Clifford transformations on graph states \cite{VDM04}:
the successive application of a simple graph transformation
rule, the so called {\em local complementation}, on a certain
graph $G$ and on those that are obtained from $G$, allows us to
generate the whole LC class of $|G\rangle$. The entire set of
graphs connected to a given $G$ by a sequence of local
complementations is usually referred to as the {\em orbit} of
$G$ (i.e., the LC class of $|G\rangle$).

Any of the graph states belonging to a given LC class could be
used as a representative for that orbit, but there is a
practical advantage in taking one requiring: (a) the minimum
number of controlled-$Z$ gates, or (b) the minimum preparation
depth of the class. If we can identify which LC class a given
$|G\rangle$ belongs to, then we can prepare $|G\rangle$ by
preparing instead the LC-equivalent state $|G'\rangle$
requiring (a) and (b), and then transform $|G'\rangle$ into
$|G\rangle$ by means of single-qubit Clifford operations
corresponding to a sequence of local complementations carried
out on $G'$. This last transformation from $|G'\rangle$ to
$|G\rangle$ requires only one additional time step, since the
local complementation at vertex $a$ is equivalent \cite{HEB04,
HDERVB06} to the unitary operation
\begin{equation} \label{LCmath}
U_a^\tau (G) = \exp \left(-i\frac{\pi}{4} \sigma_x^{(a)}\right) \prod_{b \in \mathcal{N}(a)}\exp \left(i \frac{\pi}{4} \sigma_z^{(b)} \right),
\end{equation}
where $\sigma_x^{(a)}$ is the Pauli matrix $\sigma_x$ acting on
qubit $a$, $\sigma_z^{(b)}$ is the Pauli matrix $\sigma_z$
acting on qubit $b$, and $ \mathcal{N}(a)$ is the neighborhood
of $a$. Operations on different qubits commute, and therefore
one can group together a sequence of local complementations
$U_a^\tau (G)$ into $n$ one-qubit gates $R_i$ (with
$i=0,\ldots,n-1$), which can be jointly performed in a single
step.

The preparation procedure of a graph state $|G\rangle$ through
the optimum LC-representative provides an advantage in time
steps when compared to the standard graph-based controlled-$Z$
procedure when
\begin{equation}\label{advantage}
\chi'(G)-\chi'(G')>1,
\end{equation}
since the preparation depth for the standard procedure is
$\chi'(G)$, while the preparation depth through the optimum
LC-representative is $\chi'(G')+1$, that is, the sum of the
preparation depth of the optimum LC-representative plus a unit
of time corresponding to the one-qubit gates. This means that
the preparation depth through the optimum LC-representative
provides an advantage for most graph states. For instance, as
we will see, for the graph state of Fig.~\ref{grafo-complex8},
the optimum preparation circuit requires only seven
controlled-$Z$ gates and three time steps: saving six
controlled-$Z$ gates and requiring four time steps fewer than
in the standard procedure.

However, the preparation through the optimum LC-representative
requires us to identify which LC class $|G\rangle$ belongs to.
To be practical, this must require us to identify the simplest
signature of the class.

In Sec.~\ref{classLC}, we classify all LC classes for graph
states up to $n=12$ qubits. This classification is based on a
reduced set of invariants which allows us to identify which
class a given state belongs to. In Sec.~\ref{optG}, we provide
a representative of the class with both properties (a) and (b),
if it exists, or, if it does not, one with property (a) and one
with property (b). All these results, which occupy several
hundreds of megabytes, are organized in two tables, one for
$n<12$ and one for $n=12$, and presented as supplementary
material \cite{supp1}. Finally, in Sec.~\ref{addLC}, we explain
how to obtain the one-qubit gates needed, and provide as
supplementary material a computer program \cite{supp2} to,
given the graph $G$ corresponding to the state we want to
prepare, generate a sequence of local complementations which
connect $G$ to the corresponding optimum graph(s). In
Sec.~\ref{ex}, the whole method is illustrated with an example.


\section{Classification of graph states in terms of entanglement}
\label{classLC}


The classification of the entanglement of graph states has been
achieved up to $n=7$ \cite{HEB04, HDERVB06}, and has recently
been extended to $n=8$ \cite{CLMP09-1}. There are 45 LC classes
for graph states up to seven qubits, and $101$ LC classes for
eight-qubit graph states, which are ordered according to
certain criteria based on several entanglement measures, which
are invariant under LC transformations.

Here we extend the classification up to $n=12$. The number of
orbits for $n \le 8$ qubits is $146$. For $n=9$, there are 440
orbits; for $n=10$, there are 3\;132 orbits; for $n=11$, there
are 40\;457; and, for $n=12$, there are 1\;274\;068. All the
information about each LC class is presented as supplementary
material \cite{supp1}.

For each LC class we give the number of nonisomorphic graphs in
that class (size of the orbit), $|LC|$. Then, the orbits are
classified according to the number of vertices (qubits), $|V|$;
the minimum number of edges of a graph belonging to the class
(controlled-$Z$ gates needed for the preparation), $|E|$; the
Schmidt measure, $E_S$ (upper and lower bounds are given where
the exact value is unknown); and for $n/2 \ge i \ge 2$, the
rank indexes $RI_i$ for bipartite splits with $i$ vertices in
the smallest partition. The classifying labels $|V|$, $|E|$,
$E_S$, and $RI_i$ are applied in this order. Additionally, we
include the information regarding the existence (or not) of a
two-colorable graph belonging to the class (a piece of data
which is useful, in some cases, to calculate lower and upper
bounds for the Schmidt measure).

However, these numbers, which were (almost) enough to identify
every class if $n \le 8$, are not enough to identify every
class if $n>8$. In other words, the set of entanglement
measures for $n$-qubit graph states used in \cite{HEB04,
HDERVB06, CLMP09-1} failed to distinguish between inequivalent
classes under local Clifford operations if $n > 8$: different
LC classes had coincident values for those entanglement
measures. In fact, the number of problematic LC classes
increases with $n$. Therefore, using these invariants for
deciding which entanglement class a given state belongs to is
unreliable. A finite set of invariants that characterizes all
classes has been proposed in \cite{VDD05}. However, already for
$n=7$, this set has more than $2 \times 10^{36}$ invariants
which are not explicitly calculated anywhere, and hence this
set is not useful for classifying a given graph state.

Nevertheless, a compact set of invariants related to those
proposed in \cite{VDD05} is enough to distinguish among all
inequivalent LC classes with $n \le 8$ qubits: the 4 two-index
invariants called cardinalities-multiplicities \cite{CLMP09-2}.
There is a straightforward procedure for calculating these four
invariants using the information contained in the graphs.

We have analyzed the utility and limitations of the
cardinalities-multiplicities (C-M hereafter) as LC-class
discriminants for graph states up to $12$ qubits, a question
that was left as an open problem in \cite{CLMP09-2}, where it
was conjectured that four of these C-M invariants would be
enough to label and discriminate all the LC classes. Our
results show that for graph states of $n \ge 9$ qubits the C-M
invariants fail to distinguish between inequivalent LC classes:
the smallest counterexample of the conjecture corresponds to a
pair of nine-qubit orbits that have exactly the same entire
list of C-M invariants. These are the only problematic orbits
for $n=9$ qubits. As an alternative for discriminating between
them, we have calculated the whole list of Van den
Nest-Dehaene-De Moor (VDD) invariants of type $r=1$
\cite{VDD05} for these two orbits, and once again these
invariants coincide. In order to determine the number of C-M
and VDD ``problematic'' (undistinguishable) orbits, we have
extended our calculations up to $n=12$ qubits. The ratio
$p_{f}(n)$ of the number of graphs belonging to problematic
orbits and the overall number of graphs for each $n$, gives the
probability that a randomly chosen graph state falls in one of
the problematic orbits. The values of $p_{f}(n)$ are,
fortunately, quite low (see Table~\ref{Table-pf}). Therefore,
the first step of the procedure of preparation, i.e., the
identification of the orbit, resorts to C-M invariants (and,
sometimes, type $r=1$ VDD invariants), and works in most cases.
For those rare states whose orbit identification through C-M or
VDD ($r=1$) invariants is not univocal, one would resort to VDD
invariants of higher order $r$ \cite{VDD05}. However, the
computational effort of this task makes this procedure less
efficient than simply generating the whole LC orbit of the
graph.


\begin{table}[h]
\caption{\label{Table-pf}Orbits for which C-M and VDD invariants are not good LC discriminants.}
\begin{ruledtabular}
{\begin{tabular}{cccc} $n$ & No.\ of orbits & No.\ of problematic orbits & $p_{f}(n)$ \\
\hline \hline
$\le 8$ & $146$ & $0$ & $0$ \\
$9$ & $440$ & $2$ & $0.0012218$ \\
$10$ & $3\;132$ & $8$ & $0.0006996$ \\
$11$ & $40\;457$ & $78$ & $0.0011929$ \\
$12$ & $1\;274\;068$ & $472$ & $0.0000949$ \\
\end{tabular}}
\end{ruledtabular}
\end{table}


In the supplementary material \cite{supp1}, those LC classes
which have the same set of quantities $\{|V|, |E|, E_S, RI_i\}$
are ordered according to the C-M invariants, going from the
smallest cardinality (zero) to the biggest one, and increasing
the value of the associated multiplicity of each cardinality.
For those orbits with the same set of quantities $\{|V|, |E|,
E_S, RI_i, \mbox{C-M invariants}\}$, orbits are ordered by the
increasing size of the orbit, $|LC|$. We do not apply any
subsequent classifying criteria, in case there were
undistinguishable LC classes left.

C-M invariants are given as an ordered list of multiplicities
$M_i$, for $i = 0,\ldots,x$. The value $M_i$ is the
multiplicity of the cardinality $i$. We do not list all $2^n$
possible multiplicities, but only the multiplicities of
cardinalities $0,\ldots,x$, where $x$ is the smallest number
such that all``non-problematic'' orbits are distinguished. This
may not be the smallest possible set of C-M invariants. For
instance, only the cardinalities $\{0,1,3,4\}$ are needed for
$n \le 8$, but, for simplicity's sake, we have included the
continuous list $\{0,1,2,3,4\}$.


\section{Optimum representative}
\label{optG}


We have determined the optimum representatives for all the
orbits up to $12$ qubits, initially defined as one of those
with the minimum number of edges in the orbit and, among them,
one of those with the minimum chromatic index. In the
supplementary material \cite{supp1}, we have included a column
labeled $\min(|E|, \chi', \#)$ for each LC class, where $|E|$
is the minimum number of edges in the class, $\chi'$ is the
minimum chromatic index of the graphs with $|E|$ edges, and
$\#$ is the number of nonisomorphic graphs with $|E|$ edges and
chromatic index $\chi'$. The value of $\chi'$ coincides with
the preparation depth of those representative states and
indicates how much more efficient the preparation procedure we
are proposing is than the standard one [it is $\chi'(G')$ in
Eq.~\eqref{advantage}].

\begin{figure*}[h]
\centerline{\includegraphics[width=1.8\columnwidth]{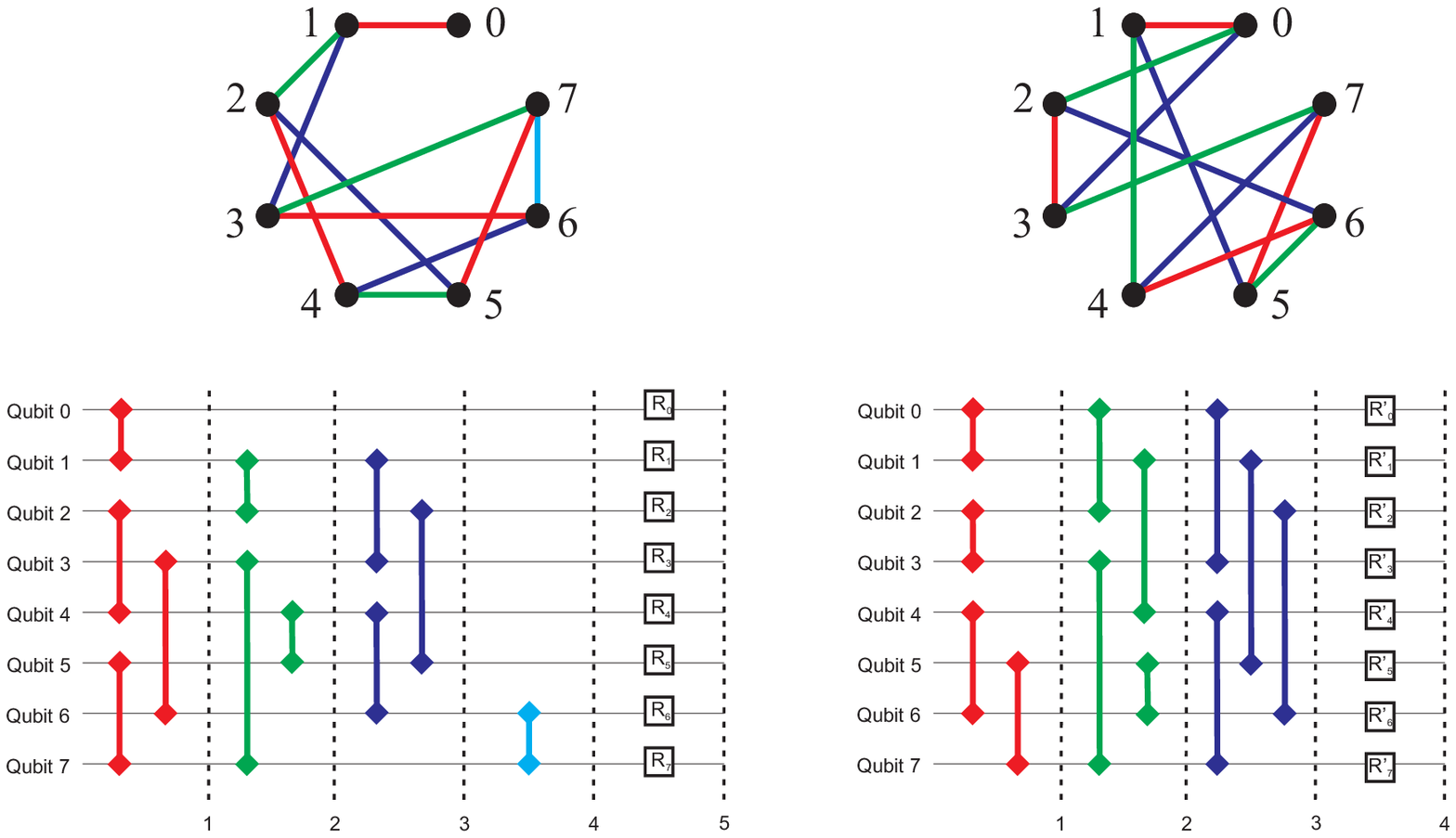}}
\caption{\label{LC136} Two optimum representatives for LC class
number 136 (up). The graph on the left is obtained by applying
over the entire orbit the filter $\min(|E|, \chi')$, i.e.,
first minimizing over $|E|$ and then over $\chi'$, whereas the
one on the right is obtained by applying the filter
$\min(\chi', |E|)$. Any graph state belonging to LC class
number 136 could be prepared by means of the circuits depicted
under those representatives. The circuit on the left
prioritizes a lower number of entangling gates; the one on the
right prioritizes a lower preparation depth. $R_i$ and $R'_i$
are one-qubit gates. They are specific for the state of the
class 136 we want to prepare.}
\end{figure*}

While carrying out the LC classification, an interesting
observation arose: the optimum representative of a certain
orbit was determined by the application of two filters to the
orbit {\em in a certain order}. First, we looked for the graph
with the minimum number of edges (which implies a minimum
number of two-qubit entangling operations), $|E|$, and second
the minimum chromatic index $\chi'$ fixed that $|E|$ (which
means a minimum preparation depth given those graphs with $|E|$
edges). However, if we commute the order of the filters, for $n
\le 7$ qubits, the result of the permutation of the filters
gives the same graph, but this is not the case for $n>7$. We
get the simplest example of non coincidence between the final
optimum representatives for $n=8$, where there is only one
orbit (LC class number 136) in which the permutation of the
filters does not produce the same final graph. There are two
graphs in this orbit with $|E|=11$ and $\chi'=4$, and one graph
with $\chi'=3$ and $|E|=12$ (see Fig.~\ref{LC136}). For each $n
\le 12$, we calculated the number of orbits with different
final representatives when we applied the filters in different
order: the ratio of the number of ``exceptional'' orbits and
the entire number of orbits increases with $n$, for $n \ge 9$
(Table \ref{Table-filter}).


\begin{table}[h]
\caption{\label{Table-filter}Orbits for which a different order
of the filters relating the minimum number of edges and minimum
chromatic index produces a non-coincident representative
graph.}
\begin{ruledtabular}
{\begin{tabular}{ccc} $n$ & $\#$ of orbits & $\#$ of exceptional orbits \\
\hline \hline
$\le 7$ & $45$ & $0$ \\
$8$ & $101$ & $1$ \\
$9$ & $440$ & $3$ \\
$10$ & $3\;132$ & $65$ \\
$11$ & $40\;457$ & $2\;587$ \\
$12$ & $1\;274\;068$ & $136\;518$ \\
\end{tabular}}
\end{ruledtabular}
\end{table}


Moreover, we have included in our tables (see supplementary
material \cite{supp1}), beside the column labeled $\min(|E|,
\chi', \#)$, another column labeled $\min(\chi', |E|, \#)$,
with a $3$-tuple of values $(\chi', |E|, \#)$, where $\chi'$ is
now the minimum chromatic index in the class, $|E|$ is the
minimum number of edges of the graphs with chromatic index
$\chi'$, and $\#$ is the number of nonisomorphic graphs with
chromatic index $\chi'$ and $|E|$ edges. Checking the
coincidence of $\chi'$ and $|E|$ in both columns directly tells
us if a certain orbit is exceptional or not. In case of
coincidence, we have left the second column blank. From an
experimental point of view, the appropriate order for the
filters is something that the experimentalists should
elucidate, since it is related to the physical substrate used
to implement the qubits, and the resources at their disposal.
If minimizing the number of two-qubit entangling operations is
a critical factor, because the fidelity in performing such
quantum gates is lower than desirable, then the appropriate
order is $(|E|,\chi')$: once the number of gates is minimized,
then it is the turn of reducing the preparation depth.


To complete our results in \cite{supp1} we provide, in the two
final columns for each LC class, an optimum graph resulting
from the filters applied in the order $\min(|E|, \chi', \#)$,
and another one applying the filters as $\min(\chi', |E|, \#)$
(if it is not coincident with the former; in case of
coincidence, the second final column is left blank). The edges
of the graph are listed, with vertices indexed as $0, \ldots,
n-1$. Moreover, edges are divided in classes (enclosed by
parentheses) that define a proper edge-coloring (with $\chi'$
colors). This information is equivalent to providing an {\em
optimum circuit} [with a number of time steps equal to
$\chi'(G')$] for preparing the optimum representative graph
$G'$ of the class. The graphs and circuits in Fig.~\ref{LC136}
have been designed according to the information in the
supplementary material \cite{supp1} for LC class number 136.


\section{One-qubit gates}
\label{addLC}


Assuming that an experimentalist has prepared the optimum
representative graph state $|G'\rangle$ corresponding to the
desired state $|G\rangle$, he or she needs to know at least one
sequence of local complementations connecting $|G'\rangle$ with
$|G\rangle$. The length of this LC sequence is not relevant,
due to the possibility of implementation of these local
operations as one-qubit gates in only one time step, as was
already discussed above. However, finding a way in the orbit
from $|G'\rangle$ to $|G\rangle$ is a hard task. To make the
entire orbit-based preparation procedure practical, we have
designed a computer program that accomplishes this task. Very
briefly, it uses the information about the graph $G$ related to
the state $|G\rangle$ that one wishes to obtain. The input is
the information about the edges. The program finds the optimal
graph(s) $G'$ with respect to both the number of edges and
chromatic index (these two quantities are also part of the
output), and provides a sequence of LC operations transforming
$|G'\rangle$ into $|G\rangle$. This computer program is
included as supplementary material \cite{supp2}.


\section{Example}
\label{ex}


In order to clarify the whole process, we go back to the graph
in Fig.~\ref{grafo-complex8}. As we mentioned, $G$ is a class-1
graph. Therefore, the preparation depth of $|G\rangle$ using
only controlled-$Z$ gates is 7. The orbit-based procedure
allows us to reduce significatively the preparation depth and
the number of controlled-$Z$ gates. It consists of the
following steps:

(I) To identify the orbit or LC equivalence class the graph
state $|G\rangle$ belongs to, we calculate the
cardinality-multiplicity invariants \cite{CLMP09-2}. The result
is: $\{0_{170}, 1_{35}, 3_{12}, 4_{7}\}$. Therefore, after
consulting \cite{supp1}, we conclude that the graph state
$|G\rangle$ belongs to the LC class number $68$.

\begin{figure}[t]
\includegraphics[width=0.22\textwidth]{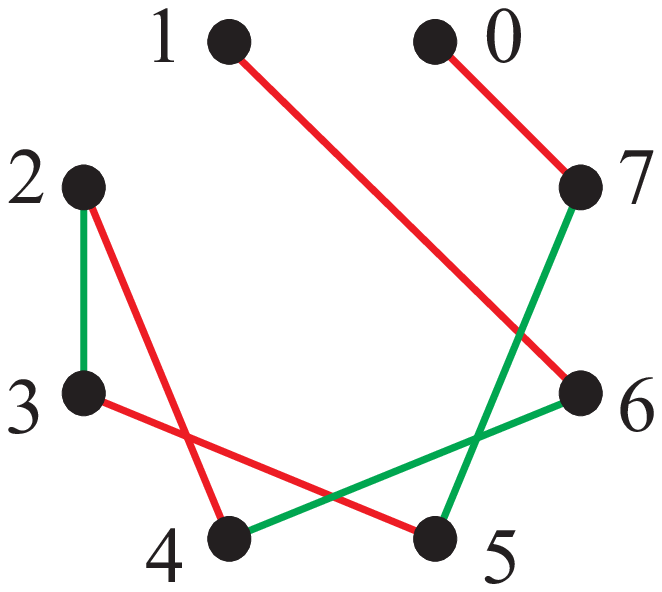}\\
\caption{Graph state $LC_8$, optimum representative of orbit 68.}\label{LC8}
\end{figure}

(II) Also in \cite{supp1} we find the optimum representative
graph $G'$: it is the eight-vertex linear cluster $LC_8$ (see
Fig.~\ref{LC8}). Graph $LC_8$ is a class-1 graph, whose maximum
degree is $\Delta(G')=2$. Hence its preparation depth is 2,
which means a remarkable saving in the preparation depth
compared to that of $|G\rangle$.


(III) Therefore, it is worth preparing $|G\rangle$ by preparing
$|G'\rangle$ and then applying suitable one-qubit gates. The
program in \cite{supp2} outputs a sequence of local
complementations which connects $G'$ to $G$. For instance, the
sequence of LC operations applied on vertices 6, 7, 4, 5, and 2
in graph $G'$ enables us to obtain $G$. Denoting the
corresponding series of local Clifford operations by $\tau
(G')$, we have $|G\rangle=\tau (G') |G'\rangle$. Applying
Eq.~\eqref{LCmath} and re-arranging terms so that $\tau
(G')=\prod_{i \in V} R_i$, where $R_i$ is a specific gate on
qubit $i$, we obtain that
\begin{subequations}
\begin{align}
&R_0 = \exp \left(i \frac{\pi}{2} \sigma_z^{(0)} \right), \\
&R_1 = \exp \left(i \frac{3 \pi}{4} \sigma_z^{(1)}\right), \\
&R_2 = \exp \left(-i \frac{\pi}{4} \sigma_x^{(2)} \right) \exp \left(i\frac{\pi}{4} \sigma_z^{(2)}\right), \\
&R_3 = \exp \left(i \frac{\pi}{2} \sigma_z^{(3)} \right), \\
&R_4 = \exp \left(i \frac{\pi}{4} \sigma_z^{(4)} \right) \exp \left(-i\frac{\pi}{4} \sigma_x^{(4)}\right) \exp \left(i \frac{\pi}{4} \sigma_z^{(4)} \right), \\
&R_5 = \exp \left(-i\frac{\pi}{4} \sigma_x^{(5)}\right) \exp \left(i \frac{\pi}{4} \sigma_z^{(5)} \right), \\
&R_6 = \exp \left(i \frac{\pi}{2} \sigma_z^{(6)} \right) \exp \left(-i\frac{\pi}{4} \sigma_x^{(6)}\right), \\
&R_7 = \exp \left(i \frac{\pi}{4} \sigma_z^{(7)} \right) \exp \left(-i\frac{\pi}{4} \sigma_x^{(7)}\right).
\end{align}
\end{subequations}

\begin{figure}[t]
\includegraphics[width=0.3\textwidth]{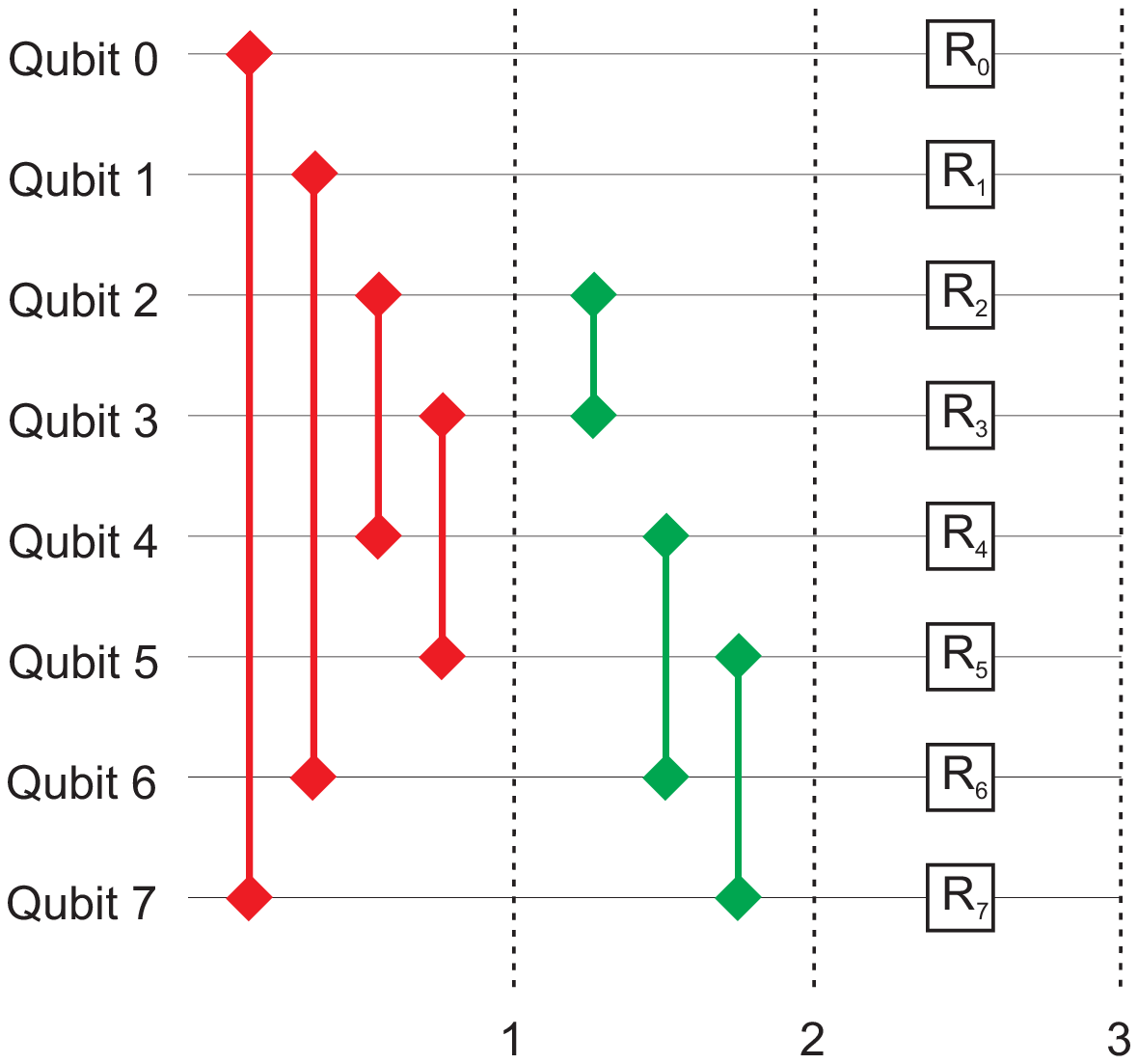}\\
\caption{Optimum circuit for preparing the graph state corresponding to Fig.~\ref{grafo-complex8}.}\label{opprep-complex8}
\end{figure}

In addition, the number of controlled-$Z$ gates necessary to
get $|G\rangle$ is remarkably reduced (six two-qubit gates
fewer than in the standard preparation method). The optimum
circuit for preparing $|G\rangle$, with a preparation depth
equal to 3, is the one in Fig.~\ref{opprep-complex8}.


\section{Conclusions}


We have proposed a procedure for the optimal preparation of any
of the more than $1.65 \times 10^{11}$ graph states with up to
12 qubits, based on their entanglement properties. Optimal
means with both (a) a minimum number of entangling gates and
(b) a minimum number of time steps, when possible, or choosing
between (a) or (b), in the other cases. The preference will
depend on the particular physical system we are considering.
The main goal has been to provide in a single package all the
tools needed to rapidly identify the entanglement class the
target state belongs to, and then easily find the corresponding
optimal circuit(s) of entangling gates, and finally the
explicit additional one-qubit gates needed to prepare the
target, starting with a pure product state and assuming only
arbitrary one-qubit gates and controlled-$Z$ gates, which
constitutes the most common scenario for practical purposes.

The results presented in this paper goes beyond those in
\cite{HEB04, HDERVB06, CLMP09-1, CLMP09-2}: the classification
of entanglement for a highly relevant family of qubit pure
states (graph states and, by extension, stabilizer states) of
9, 10, 11, and 12 qubits. In total, almost $1.3 \times 10^6$
entanglement classes are introduced.

The results have experimental relevance. For example, imagine
an experimentalist in the field of trapped ions who wants to
prepare graph states. The experimentalist knows that he can
keep, e.g., nine ions (qubits) isolated from external
influences for a given period of time, and knows that during
this time he can perform a maximum of $m$ two-qubit entangling
operations with an efficiency above a certain threshold. The
experimentalist wants to know which classes of graph states
(which classes of entanglement) are a reasonable target with
these resources. The results in this paper allow him to answer
this question: if $m=8$, he can prepare $47$ different classes
(classes 147--193 in \cite{supp1}); if $m=9$, he can prepare
$47+95=142$ different classes (classes 147--288 in
\cite{supp1}), etc. Moreover, the paper tells the
experimentalist which is the optimum sequence of lasers (gates)
required for preparing any state of each class.

More interestingly, consider that the experimentalist wants to
prepare a specific nine-qubit graph state. The paper provides
the simplest known protocol to identify which entanglement
class the target state belongs to, and give the simplest
circuit to prepare it, where simplest means in most cases the
one requiring the minimum number of entangling gates and
computational steps or, in those cases in which such a circuit
does not exist, gives a circuit requiring the minimum number of
entangling gates, and a circuit requiring minimum depth.


\section*{Acknowledgments}


The authors thank E. Kashefi and P. Moreno for useful
conversations. A.C. and A.J.L.T. were supported by MICINN
Project No.\ FIS2008-05596, A.C. was also supported by the
Wenner-Gren Foundation, L.E.D. was supported by the Research
Council of Norway, and J.R.P. was supported by MICINN Project
No.\ MTM2008-05866.


\end{document}